\documentclass[%
 aip,
 amsmath,amssymb,
reprint,%
floatfix
]{revtex4-1}

\usepackage{graphicx}
\usepackage{dcolumn}
\usepackage{bm}

\usepackage[utf8]{inputenc}
\usepackage[T1]{fontenc}
\usepackage{mathptmx}

\begin{document}

\preprint{}

\title[]{Optimization of Permalloy properties for magnetic field sensors using He$^+$ irradiation}

\author{G. Masciocchi}
\email{gmascioc@uni-mainz.de}
 \affiliation{Institute of Physics, Johannes Gutenberg University
Mainz, Staudingerweg 7, 55128 Mainz, Germany}
 \affiliation{Sensitec GmbH, Walter-Hallstein-Straße 24, 55130 Mainz, Germany}
\author{J. W. van der Jagt}%
\affiliation{ 
Spin-Ion Technologies, 10 boulevard Thomas Gobert, 91120 Palaiseau, France 
}%
\affiliation{ Université Paris-Saclay, 3 rue Juliot Curie, 91190 Gif-sur-Yvette, France}%

\author{M.-A. Syskaki}
 \affiliation{Institute of Physics, Johannes Gutenberg University
Mainz, Staudingerweg 7, 55128 Mainz, Germany}
\affiliation{Singulus Technologies AG, Hanauer Landstrasse 107, 63796 Kahl am Main, Germany}

\author{J. Langer}

\affiliation{Singulus Technologies AG, Hanauer Landstrasse 107, 63796 Kahl am Main, Germany}

\author{ G. Jakob}
 \affiliation{Institute of Physics, Johannes Gutenberg University
Mainz, Staudingerweg 7, 55128 Mainz, Germany}
\author{ J. McCord}
\affiliation{Faculty of Engineering, Institute for Material Science, Synthesis and Real Structure,
Kiel University, Kaiserstraße 2, 24143 Kiel, Germany}
\affiliation{Kiel Nano, Surface and Interface Science (KiNSIS), Kaiserstraße 2, 24143 Kiel, Germany}

\author{ B. Borie}
\affiliation{ 
Spin-Ion Technologies, 10 boulevard Thomas Gobert, 91120 Palaiseau, France 
}%

\author{ A. Kehlberger}
 \affiliation{Sensitec GmbH, Walter-Hallstein-Straße 24, 55130 Mainz, Germany}

\author{ D. Ravelosona}
\affiliation{ 
Spin-Ion Technologies, 10 boulevard Thomas Gobert, 91120 Palaiseau, France 
}
\affiliation{ 
C2N, CNRS, Université Paris-Saclay, 10 boulevard Thomas Gobert, 91120 Palaiseau, France}

\author{M. Kläui}
 \affiliation{Institute of Physics, Johannes Gutenberg University
Mainz, Staudingerweg 7, 55128 Mainz, Germany}
\date{\today}

\date{\today}

\begin{abstract}

Permalloy, despite being a widely utilized soft magnetic material, still calls for optimization in terms of magnetic softness and magnetostriction for its use in magnetoresistive sensor applications. Conventional annealing methods are often insufficient to locally achieve the desired properties for a narrow parameter range. In this study, we report a significant improvement of the magnetic softness and magnetostriction in a 30 nm Permalloy film after He$^+$ irradiation.
Compared to the as-deposited state, the irradiation treatment reduces the induced anisotropy by a factor ten and the hard axis coercivity by a factor five. In addition, the effective magnetostriction of the film is significantly reduced by a factor ten - below $1\times10^{-7}$ - after irradiation. All the above mentioned effects can be attributed to the isotropic crystallite growth of the Ni-Fe alloy and to the intermixing at the magnetic layer interfaces under light ion irradiation. We support our findings with X-ray diffraction analysis of the textured Ni$_{81}$Fe$_{19}$ alloy. Importantly, the sizable magnetoresistance is preserved after the irradiation. 
Our results show that compared to traditional annealing methods, the use of He$^+$ irradiation leads to significant improvements in the magnetic softness and reduces strain cross sensitivity in Permalloy films required for 3D positioning and compass applications. These improvements, in combination with the local nature of the irradiation process make our finding valuable for the optimization of monolithic integrated sensors, where classic annealing methods cannot be applied due to complex interplay within the components in the device. 

\end{abstract}

\maketitle



\section{Introduction}



Permalloy, a typical soft magnetic Ni-Fe alloy is employed as an active sense layer in several magnetoresistive (MR) sensor applications\cite{khan2021magnetic}. To have a small magnetostriction and low coercivity, most of these devices are designed around the alloy composition of Ni$_{81}$Fe$_{19}$ which also possesses significant anisotropic magnetoresistance (AMR). Optimization of Permalloy for AMR sensors has been studied for a long time \cite{kwiatkowski1986permalloy,groenland1992permalloy, trutzschler2014optimization} and includes different aspects:  primarily, improvement of magnetic softness and low magnetostriction.
To reach that, negligible crystalline anisotropy is firstly required. The single thin film elements typically feature a stripe shaped geometry to induce a strong shape anisotropy, providing the sensor with a well-defined  orientation of sensitivity. Furthermore, this design of the sensitive elements ensures a fixed configuration of the magnetic domains, thus enabling a very high signal-to-noise ratio. Additional anisotropies of other sources, if not oriented in the same direction as the shape anisotropy, would hinder this sensitivity direction \cite{jogschies2015recent}. Moreover, to achieve low hysteresis, the coercivity in the hard axis magnetization direction must be very low and the specific AMR  must be as high as  possible\cite{lenssen2000robust} to maximize sensitivity. Eventually, to avoid parasitic anisotropies, low magnetostriction (source of magnetoelastic anisotropy) is required.  In this case, strain in the material has small or negligible impact on the magnetic properties. The low magnetoelastic anisotropy is particularly important for sensors on flexible substrates \cite{khan2021magnetic,canon2021magnetosensitive, oliveros2021printable, melzer2019review, amara2018high} that have attracted great attention in recent years in wearable electronics and biomedical applications. To obtain this particular material property, growth optimization\cite{wang2012observation} and annealing\cite{iwata1969annealing} are viable options. However, none of these technique allow for a local treatment of the film. 

It is well known that ion irradiation is an excellent tool to tune locally the magnetic and structural properties of thin films through ordering\cite{jiang2018improving,jaafar2011pattern, devolder2013irradiation, ravelosona2000chemical} and interface intermixing\cite{masciocchi2022control,de2022local,zhao2019enhancing,diez2019enhancement}. In Permalloy films, ion irradiation has been shown to change the magnetic anisotropy\cite{woods2002local, schindler1964effect, mougin2001local} and the magneto-resistive response in the presence of exchange bias\cite{trutzschler2014optimization, trutzschler2016magnetic}. However, most of these works use ion implantation \cite{fassbender2006structural,fassbender2006control,fassbender2006mixing} or heavy ions\cite{gupta2008argon}, which can result in significant damage to the sample. This can be avoided by using lighter ions - like He$^+$ - with energies in the range of 10-30 keV \cite{fassbender2004tailoring,masciocchi2022control}. In this way, collision cascades are absent and the structural modifications are confined to the vicinity of the ion path in a metal.
Furthermore, the effect of irradiation on the magneto-elastic properties of single Permalloy films and a direct comparison between field free ion irradiation and annealing has not yet been reported\cite{baglin1997effects}.

In this work, we propose and explore the use of He$^+$ ion irradiation on sputtered layer of Ni$_{81}$Fe$_{19}$(30 nm) as material preparation for magnetic field sensors and we compare it with standard field free annealing.  Using Kerr microscopy and Vibrating Sample Magnetometry (VSM) we show that 20 keV He$^+$  ions significantly reduce the coercivity and the induced magnetic anisotropy of our magnetic material. The result is a soft magnetic film with in-plane magnetic anisotropy $<10$ J/m$^3$  and coercive field $\simeq 0.05$ mT, which is a further improvement over the values that can be obtained by field-free annealing process by a factor 5 and 10, respectively.
The anisotropy measurements are supported by a detailed comparison using the remanent domain pattern.  Additionally, we show that the polycrystalline magnetostriction can be progressively reduced by a factor ten for irradiation doses of $5\times10^{16}$ cm$^{-2}$. This reduction in magnetoelastic coupling is attributed to crystallization and changes to the interface magnetostriction caused by intermixing at the magnetic layer boundaries.  We support our findings with structural characterization performed using X-ray diffraction (XRD). The results show an overall improvement in the crystallization after irradiation and annealing. We attribute the reduction in magnetic anisotropy to the absence of a preferential direction of atomic ordering and to stress relaxation during irradiation. 
As post growth He$^+$ ion irradiation improves magnetic softness and minimizes strain cross sensitivity of Permalloy, AMR magnetic sensors with high sensitivity and low hysteresis can be envisioned even for integrated devices.


\section{Experimental methods}

The samples have been prepared by DC magnetron sputtering using a Singulus Rotaris system on a 1.5 $\mu$m thick, thermally oxidized SiOx on top of a 625 $\mu$m thick $Si$ substrate. A layer of Ni$_{81}$Fe$_{19}$ (30 nm) is sputtered at room temperature in the presence of a rotating magnetic field of 5 mT on a NiFeCr (5 nm) seed layer and capped with 4 nm of Ta as shown in Fig. \ref{fig_3} (b). The following sputtering conditions were used for the magnetic layer growth: base pressure $5\times10^{-8}$ mbar,  sputtering power 1200 W and Ar$^+$ flow 90 sccm. The seed layer is used to promote a NiFe (111) texture during growth and it is known to improve magnetoresistance\cite{fassbender2006structural}.
After deposition, optical lithography and ion etching have been  used to pattern arrays of disks (80 $\mu$m of diameter and 3 $\mu$m of spacing) on the samples in order to probe the local film properties. Multiple copies of the samples have been irradiated at an energy of 20 keV with different fluences of He$^+$ ions from $5\times 10^{13} $ to $5\times 10^{16} $ cm$^{-2}$. At these irradiation conditions, the majority of the ions reach the substrate (roughly 94$\%$ from Monte Carlo TRIM \cite{ziegler2010srim} simulations, not shown), resulting in homogeneous irradiation of the entire layer stack. 

To compare the effect of ion irradiation to thermal annealing, the same magnetic material has been consecutively annealed for three hours at 200, 265 and 300$^\circ$C at a pressure of 10$^{-7}$ mbar. In order to avoid a magnetization induced preferential direction of ordering\cite{fassbender2006control,okay2018tailoring}, external magnetic fields have been minimized during the irradiation and annealing steps. The thin film magnetic properties have been measured with Kerr microscopy and VSM. The magnetic properties of our films are summarized in Table \ref{tab_material_film}. Due to the negligible implantation\cite{fassbender2004tailoring}, the value of the Young's modulus is assumed to be unaffected by our irradiation and annealing step. Electrical measurement of anisotropic magnetoresistance (AMR) have been performed with four contacts in line in the presence of a rotating magnetic field of 10 mT.

\begin{table}[h!]
    \centering
    \begin{tabular}{||c c c c c c||} 
 \hline
    $Ni_{81} Fe_{19}$ & $M_{s}$ (T) & $K_{u}$ (J/m$^3$)  &  $H_c$ (mT)  & $\lambda_s$ x$10^{-6}$ & $Y$  (GPa) \\ [0.5ex] 
 \hline\hline

 as-deposited & 0.95(1) &	78(5) &	0.20(5) &	-0.7(1) & 200\cite{klokholm1981saturation} \\ 
 
 \hline
  Ann. 265$^\circ$C & 0.95(1) &	70(5) &	0.15(5) &	+0.04(9) & 200\cite{klokholm1981saturation} \\
   \hline
     He$^+$ $5\times10^{16}$ cm$^{-2}$ & 0.91(1) &	8(7) &	0.05(5) &	+0.01(9) &  200\cite{klokholm1981saturation}\\
   \hline
\end{tabular}
\caption{Parameters of the magnetic materials (thickness 30 nm) after deposition, annealing and He$^+$ ion irradiation. The values without reference are quantified experimentally. Here, $M_s$ is the saturation magnetization, $K_u$ is the uniaxial anisotropy constant, $H_c$ is the coercive field, $\lambda_s$ is the saturation magnetostriction and $Y$ is the Young's modulus. The same value for $Y$ is considered in all cases. }
\label{tab_material_film}
\end{table}

To apply strain to our devices, the substrate was bent mechanically with a three-point bending method. As reported in our previous work \cite{masciocchi2021strain} a tensile and uniaxial strain is generated\cite{raghunathan2009comparison}. Moreover the strain is uniform in the central area of the sample and thus in the measured region. As the thin films are in total 40 nm thick, we assume that the strain is entirely transferred from the substrate and that shear strain is negligible. 
Structural modifications caused by ion irradiation and annealing were probed by X-Ray Diffraction (XRD) using a Bruker D8 Discover system. Angular $2\Theta/\Theta$ scans and rocking curve measurements were performed on 1 by 1 cm samples. 


\section{Results and discussion}



\begin{figure}[h!]
\centering\includegraphics[width=9cm]{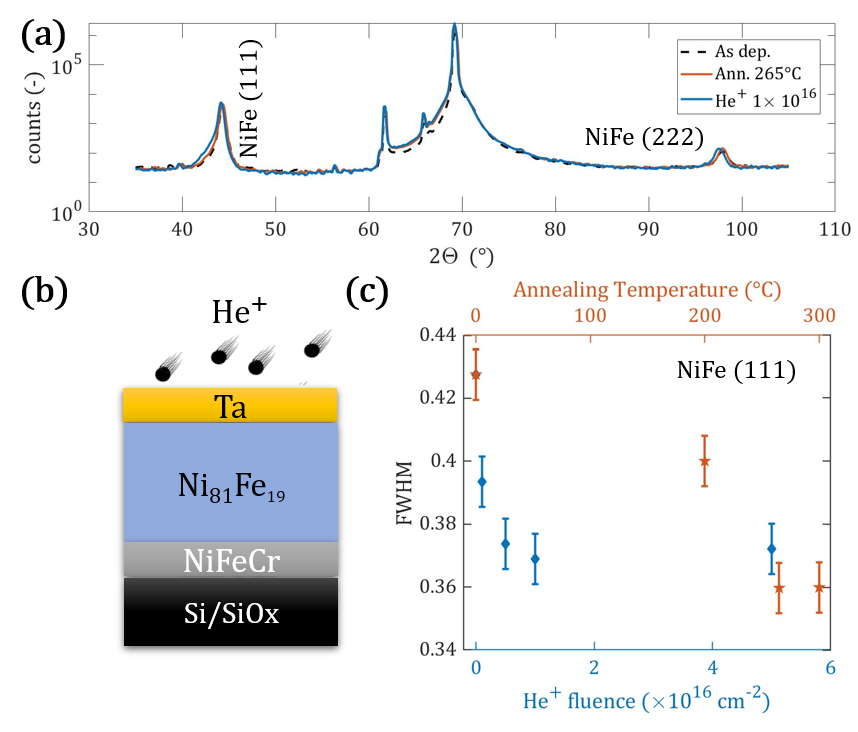}
\caption{\label{fig_3} (a) $2\Theta/\Theta$ XRD angular scan of the NiFe samples  for the sample in the as-deposited state, after annealing and after irradiation. (b) schematic of the NiFeCr(5 nm)/Ni$_{81}$Fe$_{19}$(30 nm)/Ta(4 nm) stack. (c) FWHM of the NiFe (111) peak as a function of He$^+$ fluence  and as function of annealing temperature.}
\end{figure}

To compare the structural modifications induced by different material treatment on a Ni-Fe alloy, XRD measurements on the Ni$_{81}$Fe$_{19}$ (30 nm) film as-deposited and after irradiation and annealing are performed and reported in Fig. \ref{fig_3}. Fig. \ref{fig_3} (a) shows $2\Theta/\Theta$ angular scan of the Permalloy film. A well defined crystalline texture of NiFe (111) (and its second order peak) is present for the material in the as-deposited state and persists after irradiation and annealing in all the fluence and temperature range explored. The full width at half maximum (FWHM) of the (111) peak is reported in Fig. \ref{fig_3} (c) as a function of the irradiation fluence (blue diamonds) and the temperature during annealing (orange pentagrams). In both cases, the FWHM of the (111) peak decreases by about 15$\%$ with increasing ion fluence and annealing temperature with respect to the as-deposited case.  The crystallite size (or the size of a coherently diffracting domain in the material) is a fundamental property that can be extracted from XRD profile\cite{ahmadipour2016assessment}. According to the Scherrer equation\cite{zak2011x},

\begin{equation} \label{eq_scherrer}
D=\frac{K \lambda}{\beta cos\theta}
\end{equation}

the size of crystallites is inversely proportional to the FWHM of a diffraction peak. Here $K=0.9$ is a dimensionless shape factor, D the crystallite size, $\lambda$ the wavelength of the Cu-K$\alpha$ radiation, $\theta$ the diffraction angle and $\beta$ is the line broadening at FWHM of the XRD peak in radians, after subtracting the instrumental line broadening. As our measurements show, both annealing at 265$^\circ$C and ion irradiation with a fluence of $5\times10^{16}$ cm$^{-2}$ increase the size of crystallites in our films. The estimated size of the diffracting domains using eq. \ref{eq_scherrer} is 22(1) nm for the as-deposited case and 24(1) nm after the two material treatments.  Additionally, rocking curve measurements of the NiFe (111)  peak were preformed and more information can be found in section S1 of the supplementary material. Both for the irradiated samples and for the annealed ones a decrease in the FWHM of the rocking curve is observed indicating improvement in the film crystalline phase\cite{park2005ion}. 

The major effect of room temperature irradiation has been shown to be improved material uniformity\cite{van2022revealing} and interface intermixing\cite{masciocchi2022control}. In the same way, thermal annealing is widely used to induce crystallization\cite{wang2009situ} and promote atomic diffusion\cite{schulz2021increase}. Similar effects have been observed in literature for amorphous alloys, where annealing \cite{mccord2005magnetic} and He$^+$ irradiation\cite{devolder2013irradiation,ravelosona2000chemical} providing high short range atomic mobility allow a mechanism for growth of the ordered phase at the expense of its disordered or less ordered counterpart.

\begin{figure}[h!]
\centering\includegraphics[width=9cm]{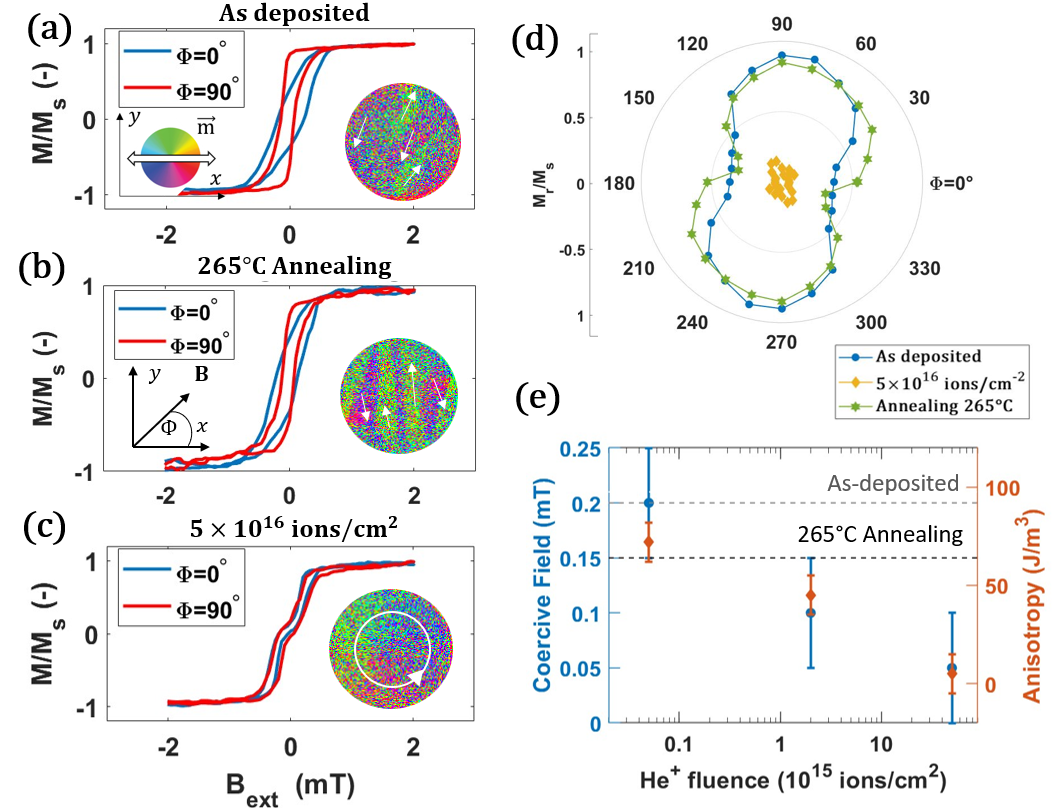}
\caption{\label{fig_1} (a) - (c) in-plane hysteresis loops of NiFeCr(5 nm)/Ni$_{81}$Fe$_{19}$(30 nm)/Ta(4 nm) after sputtering, after thermal annealing and after He$^+$ ion irradiation, respectively. In the inset, the corresponding remanent magnetic state ($B_{ext}=0$ m$T$) for 80 $\mu$m disks is shown. The field was applied along $\Phi=0^\circ$. (d) angular plot of the normalized remanent magnetization $M_r/M_s$ as function of the in-plane magnetic field direction $\Phi$ for as-deposited, irradiated and annealed samples. (e) coercive field (blue) and uniaxial magnetic anisotropy (orange) measured along the field direction $\Phi=0^\circ$ on a Permalloy sample irradiated with different fluences of ions during He$^+$ irradiation. For comparison, the values after annealing and in the as-deposited state are reported with dashed lines.}
\end{figure}

The thin film magnetic properties have been measured with Kerr microscopy and are reported in Fig. \ref{fig_1}. Figs. \ref{fig_1} (a)-(c) report the hysteresis curves for the  NiFeCr(5 nm)/Ni$_{81}$Fe$_{19}$(30 nm)/Ta(4 nm) sample for two perpendicular in-plane directions of the applied magnetic field: (a) for the as-deposited state, (b) after annealing and (c) after irradiation. The curves refer to the magnetic contrast of the structured film into 80 $\mu$m disks. 

The magnetic response of the Permalloy film in the as-deposited state can be seen in Fig. \ref{fig_1} (a). As the magnetization curves at $\Phi=0^\circ$ and $\Phi=90^\circ$ are different, a weak uniaxial magnetic anisotropy  $K_{u}$, is present in the as-deposited Ni$_{81}$Fe$_{19}$ and might be associated to internal stresses during the material growth or asymmetries in the deposition system\cite{zou2002influence}. The value of $K_{u}=80(7)$ $\frac{J}{m^3} $ has been obtained subtracting the area between the easy and hard axis loop of the as-deposited state. The direction of the magnetic easy axis anisotropy can be seen in the orientation of the magnetic domains at the remanent state (inset of Fig. \ref{fig_1} (a)). The field was applied along $\Phi=0^\circ$ and then reduced to zero. A vector image of the in-plane magnetization is obtained by the sum of the horizontal and vertical components of the magnetic contrast. In this case, the domains align along the easy axis direction.  The measurement has been repeated for the same film after annealing and is reported in Fig. \ref{fig_1} (b). After the annealing, the in-plane hysteresis loops still show the presence of uniaxial magnetic anisotropy. This is confirmed by the remanent magnetic state (inset of Fig. \ref{fig_1} (b)) as the magnetic domains again orient in the easy axis direction $\Phi\simeq90^\circ$.  Interestingly, the magnetic response  of the irradiated Permalloy reported in Fig. \ref{fig_1} (c)  is significantly different with respect to the as-deposited and annealed case. The hysteresis loops now show a negligible angular dependence on $\Phi$.  Both the magnetic anisotropy and the hard axis coercivity $H_c$ are significantly reduced. A confirmation of the extremely low magnetic anisotropy of the irradiated Permalloy can be seen in the inset of Fig. \ref{fig_1} (c). The remanent magnetic configuration is a vortex state, which is formed as  the low induced anisotropy is negligible compared to the shape anisotropy of the patterned disks.

Fig. \ref{fig_1} (d) reports the angular plot of the normalized remanent magnetization for the three samples considered. The as-deposited and the annealed case (in blue and green, respectively), show a signature of uniaxial magnetic anisotropy with easy axis and sizable remanent magnetization at $\Phi\simeq90^\circ$. The irradiated sample instead, shows reduced remanent magnetization for all the angles. The low remanent magnetization is typical for the vortex state in inset of Fig. \ref{fig_1} (c). To further understand the improvement to the magnetic softness of our Permalloy after irradiation, we have gradually increased the He$^+$ fluence (ions/cm$^2$) keeping ion energy constant. The measurements of  $H_c$ and $K_u$ as a function of the fluence of He$^+$ ions during irradiation are reported in Fig. \ref{fig_1} (e). The values of the film as-deposited and after annealing are given for comparison by dashed lines.  For low fluences, no sizable effects are noted. At fluences larger than $5\times10^{13}$ cm$^{-2}$ the coercivity and the anisotropy are progressively reduced as the He$^+$ fluence is increased. For the maximum fluence of $5\times10^{16}$ cm$^{-2}$, $H_c$ is five times lower compared to the as-deposited state while the induced anisotropy is decreased by a factor ten. We do not observe a similar substantial reduction of these magnetic parameters after the annealing.    

A possible explanation for this dissimilarity is the different mechanism of ordering promoted during irradiation and field-free annealing.
Improved atomic ordering in Permalloy after annealing and irradiation with different ions\cite{gupta2008argon,gupta2005influence} has been reported in literature. Some of these studies on polycrystalline films\cite{jian1989ar+,liu1987ion} show that 
 crystalline grain growth is more homogeneous  for irradiation than for thermal annealing in the temperature range from 200 to 300$^\circ$ C. This difference originates from the distinct mechanism with which chemical ordering of the alloy is changed during the two processes\cite{liu1987ionn}. 
 As we see from these studies, radiation-enhanced mobility is more isotropic in the absence of an applied magnetic field, if compared to heat-induced mobility\cite{schindler1964effect,liu1987ionn}. Accordingly, a stronger reduction in the magnetic anisotropy for the irradiated samples can be expected.



Recently\cite{van2022revealing}, a comparison between ion irradiation and thermal annealing  analyzing the microscopic pinning parameters for DW motion has been conducted. In this work, the annealed sample shows strong but widely distributed pinning sites. In contrast to this, the irradiated sample exhibits weaker defects with a higher density.  A possible explanation for the observed reduction in coercivity in our irradiated samples, is therefore  an overall smoother DW energy  landscape after irradiation, which allows for domain formation and switching of the magnetization at lower magnetic fields. In addition to that, the release of internal stresses in the film, that has been reported during irradiation\cite{devolder2000light, maziewski2012tailoring}, can also be responsible for improvements to the soft magnetic properties of our Permalloy\cite{zou2002influence}. 

\begin{figure}[h!]
\centering\includegraphics[width=9cm]{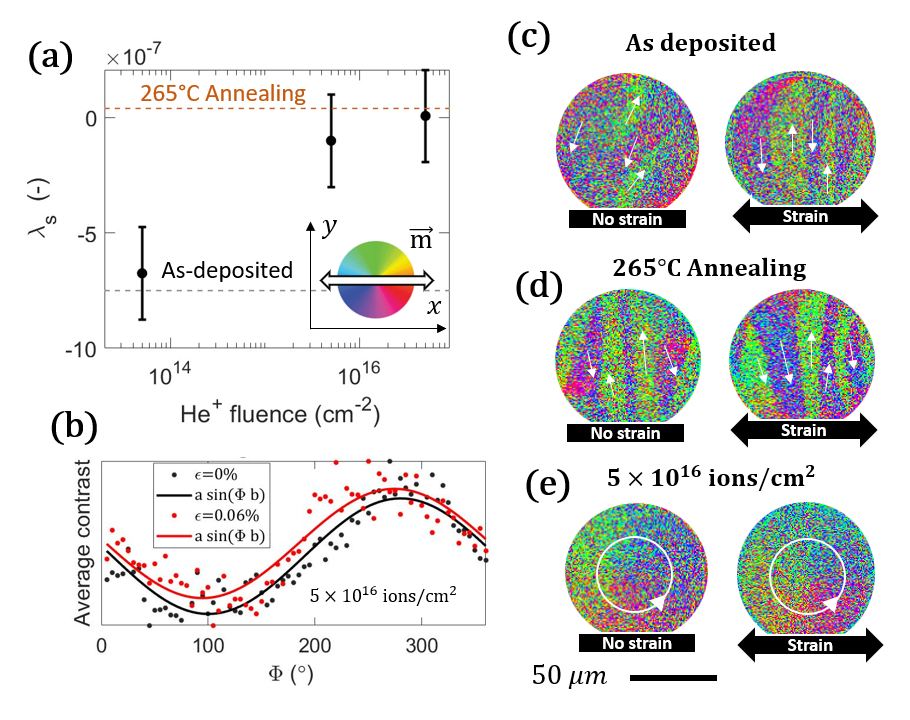}
\caption{\label{fig_2} (a) saturation magnetostriction $\lambda_s$ as a function of He$^+$ ions fluence during irradiation. The values as-deposited and after annealing are reported for comparison with dashed lines. (b) average contrast for 80 $\mu$m disks as a function of the in plane angle $\Phi$ for the irradiated sample in the remanent state (magnetic vortex state) before and after the application of strain.  (c) - (e) remanent magnetic state for  80$\mu$m diameter disks before (left) and during (right) uniaxial strain $0.06\%$ application for as-deposited, annealed and irradiated Permalloy, respectively.  }
\end{figure}

 To evaluate the effect of ion irradiation and annealing on the magnetoelastic coupling of a thin magnetic Ni-Fe alloy, the strain-dependent magnetic properties have been investigated.  Uniaxial in-plane strain is applied to a full film of NiFeCr(5 $nm$)/Ni$_{81}$Fe$_{19}$(30 $nm$)/Ta(4 $nm$) by three point bending method as previously reported\cite{masciocchi2021strain}. Since the magnetization is coupled to the external strain via the expression of the anisotropy energy,  the magnetic anisotropy before and after the application of strain is measured using Kerr microscopy. A strain of $\epsilon_{xx}=0.06\%$ (tensile) is applied along the in-plane direction $\Phi=0^\circ$. The  expression for the magnetoelastic anisotropy depends on the saturation magnetostriction $\lambda_s$ of the material according to\cite{finizio2014magnetic}

 \begin{equation} \label{eq_strain_eanis}
K_{ME}=\frac{3}{2}\lambda_s Y \epsilon,
\end{equation}

where $Y$ is the Young's modulus and $\epsilon$ is the uniaxial tensile strain. Using eq. \ref{eq_strain_eanis} and the values of the Young's modulus in Table \ref{tab_material_film}, we calculate the effective magnetostriction of the film for different He$^+$ fluences. The calculated values are reported in Fig. \ref{fig_2} (a). In the as-deposited state, as well as for He$^+$ fluences in the range of $10^{13}$ cm$^{-2}$, $\lambda_s=-7(2)\times 10^{-7}$ is negative. In this case, a tensile strain increases the anisotropy field in the direction $\Phi=0^\circ$. For larger fluences of ion during irradiation, the magnetostriction is progressively reduced and reaches values close to zero for a fluence of $5\times10^{16}$ cm$^{-2}$. In this case, the magnetoelastic anisotropy  is negligible and the material is insensitive to the applied strain. For this reason, the magnetization curves before and after the application of $\epsilon_{xx}=0.06\%$ are almost unchanged. The saturation magnetostriction of the magnetic layer after annealing has been measured and is reported in Fig. \ref{fig_2} (a) for comparison. After the annealing $\lambda_s \simeq 0$ is reported.

An additional confirmation of the magnetic behavior of the stack under strain is obtained by imaging domain formation using the magneto-optical Kerr effect (MOKE). The MOKE images shown in Figs. \ref{fig_2}(c)–(e) show how the magnetoelastic anisotropy alters the preferential direction of magnetic domains before (left) and after (right) the application of strain. Let us first consider the as-deposited state (Fig. \ref{fig_2}(c)). Before the application of strain, the magnetization aligns to the deposition-induced anisotropy easy axis. After the application of strain, the negative magnetostriction of the as-deposited sample  orients the magnetic domains along the y direction, perpendicular to the uniaxial strain $\epsilon_{xx}$. Fig. \ref{fig_2}(d) shows instead the domain pattern for a sample annealed at $265^\circ$C. In this case the remanent magnetic state is almost not altered by the applied strain. This is in agreement with the extremely low magnetostriction measured, that results in negligible magnetoelastic anistropy $K_{ME}<<K_u$. The remanent state for the sample irradiated with He$^+$ fluence $5\times10^{16}$ cm$^{-2}$  (Fig. \ref{fig_2} (e)) exhibits instead a magnetic vortex state that is not altered after the application of $\epsilon_{xx}=0.06\%$. The initial vortex state, unchanged under the application of strain, highlights that the contribution of induced and magnetoelastic anisotropy have been reduced to a point that only the shape anisotropy determines the remaining domain pattern.

In order to compare more quantitatively  the MOKE  images and the vortex state of the irradiated sample, the average radial magnetization was calculated from the longitudinal component of the vector image for different in-plane $\Phi$ directions\cite{gilbert2016magnetic}. The average contrast is calculated for a single 80 $\mu$m disk for the images in Fig. \ref{fig_2}  (e) and is reported in Fig. \ref{fig_2}  (b).  For the unstrained state seen in Fig. \ref{fig_2}  (e) left, the disk’s magnetization is a circularly-symmetric vortex, and the average contrast varies periodically with angular position on the disk. The values well follow the expression $a$ $sin(\phi b)$, black line in Fig. \ref{fig_2} (b). After the application of strain, as a consequence of the extremely small magnetostriction, the average contrast, red line in Fig. \ref{fig_2} (b), still follows the periodic behavior $a$ $sin(\phi b)$. 



A possible explanation for the reported reduction in saturation magnetostriction after ion irradiation and annealing is the growth in size of the crystallites in the NiFeCr(5 nm)/Ni$_{81}$Fe$_{19}$(30 nm)/Ta(4 nm) sample, already highlighted in Fig. \ref{fig_3} (c).
The magnetostriction of isotropic oriented cubic crystallites can be written as the combination of the saturation-magnetostriction constants $\lambda_{100}$ and $\lambda_{111}$ in the (100) and (111) directions, respectively\cite{bozorth1993ferromagnetism}

 \begin{equation} \label{eq_cryst_magnetostriciton}
\lambda_s=\frac{2\lambda_{100}+3\lambda_{111}}{5}.
\end{equation}
 
 In Permalloy, the two component of the magnetostriction change significantly over the relative Ni-Fe composition range  altering the effective magnetostriction, $\lambda_s$. The composition used in this work, Ni$_{81}$Fe$_{19}$, is predicted to have $\lambda_s$ close to zero \cite{judy1994magnetic}. In our XRD measurement, a $15\%$ reduction to the  (111) peak FWHM is observed after irradiation and annealing. This crystallization can alter the relative contribution of $\lambda_{100}$ and $\lambda_{111}$ in the magnetic layer. Following eq. \ref{eq_cryst_magnetostriciton}, the effective magnetostriction of the film is changed. As shown in Fig. \ref{fig_2} (a), the magnetostriction is progressively reduced for higher fluences and annealing temperatures as the size of crystallites caused by irradiation and annealing increases. On top of that, increased intermixing at the magnetic layer boundaries, could alter the interface magnetostriction\cite{singleton2002interfacial}  (inversely proportional to the film thickness\cite{choe1999giant,song1994giant}) thus playing a role in the effective magnetostriction of the film.

To validate the usability of our Permalloy layer for sensing application, transport measurements have been conducted. The electrical characterization confirms that the NiFeCr(5 nm)/Ni$_{81}$Fe$_{19}$(30 nm)/Ta(4 nm) sample has sizable AMR $\Delta R/R=1.1(1)\%$. As the AMR does not change after irradiation, the proposed material treatment is suitable for improving magnetic properties of magnetic material for magnetic sensing applications. AMR measurements can be found in section S2 of the supplementary material.

\section{Conclusions}

In conclusion, we have investigated the effects of He$^+$ irradiation and  thermal annealing on the magnetic properties of NiFeCr(5 nm)/Ni$_{81}$Fe$_{19}$(30 nm)/Ta(4 nm). Our XRD analysis suggests that both irradiation and annealing promote crystalline growth of the textured Ni$_{81}$Fe$_{19}$ alloy. While the irradiation treatment strongly reduces the hard axis coercivity down to $0.05$ m$T$ and the deposition induced anisotropy by a factor ten, the field-free annealing does not significantly improve  the magnetic softness. We mainly attribute this to stress relaxation in the film after irradiation and to the different mechanism for atomic ordering, that is completely isotropic in the case of irradiation only. In addition, the effective magnetostriction of the film is significantly reduced by a factor ten after irradiation and annealing as confirmed by anisotropy measurement in the presence of in-plane strain. Importantly, we have shown that the sizable magnetoresistance is preserved after the irradiation.
As a result, post growth He$^+$on irradiation is an excellent tool to improve magnetic softness and minimize strain cross sensitivity of Permalloy. In contrast to thermal annealing, ion irradiation offers the advantage of performing a local material treatment\cite{woods2002local,mccord2005magnetic,fassbender2008magnetic} to adjust the anisotropy and write magnetic domain patterns directly into thin film structured devices. As a consequence, we can locally tune the properties of a magnetic material to make it suitable, for instance, for high sensitivity and low-hysteresis integrated AMR sensors that are insensitive to stain.



\section*{Supplementary Material}
See supplementary material for the rocking curve and AMR measurements.  

\begin{acknowledgments}
 This project has received funding from the European Union’s Horizon 2020 research and innovation program  under  the  Marie  Skłodowska-Curie  grant  agreement  No  860060  “Magnetism  and  the effect of Electric Field” (MagnEFi), the Deutsche Forschungsgemeinschaft (DFG, German Research Foundation) - TRR 173 - 268565370 (project A01 and B02),  the DFG funded collaborative research center (CRC)1261 / project A6  and the Austrian Research Promotion Agency (FFG). The authors acknowledge support by the chip production facilities of Sensitec GmbH (Mainz, DE), where part of this work was carried out and the Max-Planck Graduate Centre with Johannes Gutenberg University.
\end{acknowledgments}






\nocite{*}
\bibliography{bibliography}

\newpage



\end{document}


\preprint{}

\title[\textbf{Suppl. material} -  Optimization of Permalloy properties for magnetic field sensors using He$^+$ irradiation]{Supplementary material -  Optimization of Permalloy properties for magnetic field sensors using He$^+$ irradiation}

\date{\today}

\maketitle

\subsection*{\textbf{S1} - Rocking curve measurements  }

\begin{figure}[h!]
\centering\includegraphics[width=16cm]{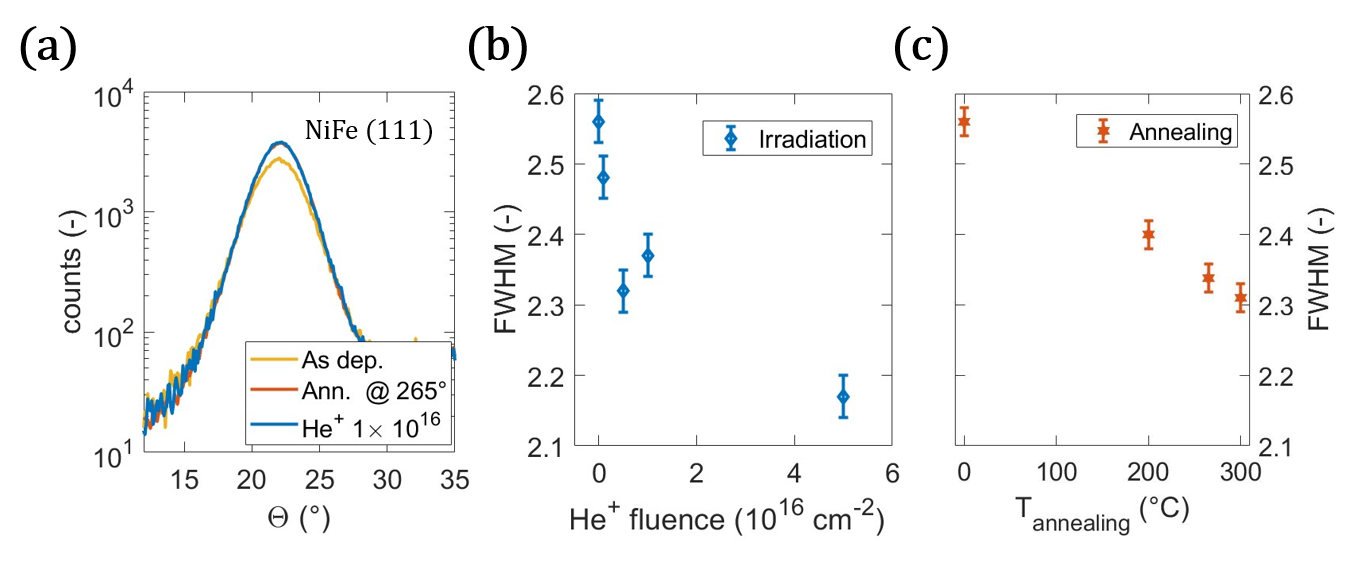}
\caption{\label{fig_S01} Details of the rocking curve measurements obtained with X-Ray diffraction. (a) rocking curve of the NiFe (111) peak for the as deposited, irradiated and annealed samples. The fitted values of the FWHM of the rocking curves are reported as a function of ion fluence during irradiation (b) and as a function of temperature during annealing (c).}
\end{figure}

Additionally, rocking curve measurements on our sample using X-Ray diffraction are performed. In a rocking curve, the detector is set at a specific Bragg angle and the sample is tilted. For this reason, the intensity is scanned along the $\Theta$ angle. A perfect crystal will produce a very sharp peak, observed only when the crystal is properly tilted so that the crystallographic direction is parallel to the diffraction vector\cite{rocking}. For this measurement, a monochromator was used. In Fig. S\ref{fig_S01} (a) rocking curve measurements obtained by X-Ray Diffraction (XRD) are presented for $1\times1$ cm$^2$ samples in the as deposited state as well as after annealing at 265$^\circ$C and after irradiation with fluence $1\times 10^{16}$ cm$^{-2}$. After irradiation and annealing, the intensity of the rocking curve increases, while the  full width at half maximum (FWHM) decreases with respect to the as deposited case (in blue).

Figs. S\ref{fig_S01} (b) and (c) contain the rocking curve FWHM as a function of dose of ions during irradiation and as a function of temperature during annealing, respectively. Both the material treatment decrease the FWHM of the rocking curve. Defects like mosaicity, dislocations, and curvature create disruptions in the perfect parallelism of atomic planes. This is observed as broadening of the rocking curve of our polycrystalline NiFe layer (111) textured. The observed reduction in FWHM is attributed to the improved crystallization, in agreement with the measurement on the (111) peak FWHM reported in the main text of this manuscript.

\subsection*{ \textbf{S2} - AMR measurements}

The  anisotropic magnetoresistance (AMR) effect occurs in 3d transition metals and can be observed macroscopically by a change of conductivity when a magnetic field is applied on a sample when current is flowing. The resistivity of the sample will depend on the angle $\Phi$ between the magnetization direction and the current flow according to \cite{AMR}
 \begin{equation} \tag{S.1}\label{eq_AMR}
R(\Phi)=R_{\perp}+(R_{\perp}+R_{\parallel})cos^2\Phi=R_{\perp}+\Delta R cos^2\Phi.
\end{equation}
The magnitude of this effect can be quantified by the magnetoresistive coefficient

 \begin{equation} \tag{S.2}\label{eq_AMR}
\frac{\Delta R}{R_{\parallel}}.
\end{equation}

\begin{figure*}[h!]
\centering\includegraphics[width=12cm]{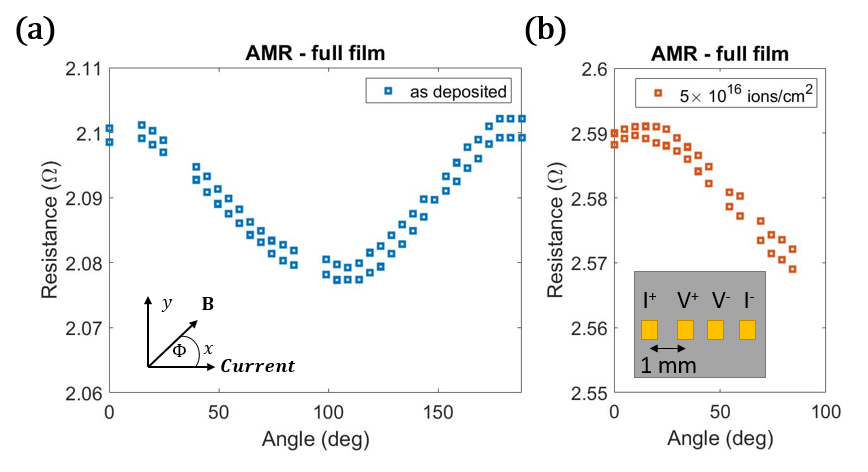}
\caption{\label{fig_S02} AMR measurements performed on a full film of Ni$_{81}$Fe$_{19}$ 30 nm before (a) and after (b) He$^+$ irradiation. A dose of $5\times 10^{16}$ cm$^-2$ is considered. In the inset, the schematic of the contacts during the AMR measurement is shown.}
\end{figure*}

To measure the AMR of our sample, we use the 4 point probe method where the current path direction is set by the external contacts. The four contacts are 1 mm apart as shown in the inset of Fig. S\ref{fig_S02}. The resistance is measured as the sample is rotated in a magnetic field of 10 mT, sufficient to saturate the magnetization along the field direction. The current is 1 $mA$.

 Fig. S\ref{fig_S02} (a) and  (b) contain the measurement of the film resistance as a function of the $\Phi$ angle for the as deposited and irradiated sample, respectively. The measured AMR is 1.1(1)$\%$ for the as deposited sample, and is not changed by the irradiation.
